\documentclass[11pt,titlepage]{article}

\usepackage{setspace}

\usepackage{amsmath}

\usepackage{graphicx}

\usepackage[round,numbers,sort&compress]{natbib}
\title{Atomic force microscopy (AFM) study of thick lamellar stacks of phospholipid bilayers}

\author{Arne~Sch\"afer, Tim~Salditt, and Maikel~C.~Rheinst\"adter\thanks{
        Corresponding author.  Present address:
        Department of Physics and Astronomy,
        University of Missouri-Columbia,
        Columbia, MO 65211, U.S.A,
        Tel.:~(573)882-3217, Fax:~(573)882-4195}\\
        Institut f\"{u}r R\"{o}ntgenphysik,\\
        Georg-August-Universit\"{a}t G\"{o}ttingen,\\
        Friedrich-Hund Platz 1,\\
        37077 G\"{o}ttingen,\\
        Germany}

\date{}

\begin{document}

\maketitle

\abstract{We report an Atomic Force Microscopy (AFM) study on thick
multi lamellar stacks of approx.\ $10~\mu$m thickness (about 1500
stacked membranes) of DMPC
(1,2-dimyristoyl-sn-glycero-3-phoshatidylcholine) deposited on
silicon wafers. These thick stacks could be stabilized for
measurements under excess water or solution. From force curves we
determine the compressional modulus $B$ and the rupture force $F_r$
of the bilayers in the gel (ripple), the fluid phase and in the
range of critical swelling close to the main transition. AFM allows
to measure the compressional modulus of stacked membrane systems and
values for $B$ compare well to values reported in the literature. We
observe pronounced ripples on the top layer in the P$_{\beta'}$
(ripple) phase and find an increasing ripple period $\Lambda_r$ when
approaching the temperature of the main phase transition into the
fluid L$_{\alpha}$ phase at about $24~^{\circ}C$. Metastable ripples
with 2$\Lambda_r$ are observed. $\Lambda_r$ also increases with
increasing osmotic pressure, i.e., for different concentrations of
polyethylene glycol (PEG).

}

\clearpage

\section*{Introduction}
There are only few techniques which allow to study structures on the
nanometer scale with molecular spatial resolution. Scattering
experiments, as, e.g., x-ray or neutron scattering, offer this high
resolution but work in reciprocal or {\em 'Fourier'} space, and the
phase problem usually inhibits a direct transformation back into
real space. The investigation of particular regions in reciprocal
space nevertheless gives valuable structural information. By
multiplying and stacking the corresponding surfaces or interfaces,
these techniques can be made surface sensitive.

Scanning probe microscopic techniques, as Atomic Force Microscopy
(AFM) \citep{Binnig} and STM (Scanning Tunnelling Microscopy) on the
other hand give high resolution real space pictures but are
essentially surface sensitive. The combination of the two techniques
is therefore a powerful tool to gain information about bulk {\em
and} surface structure. In this study, we used AFM to study the
surface structure of the uppermost layer of thick lamellar stacks of
phospholipid bilayers under excess water and solutions. As is well
known, AFM can be used to monitor biological processes under
physiological conditions \citep{Drake} on the cellular \citep{Fritz}
and the molecular scale \citep{Radmacher1994}. In addition to
topographical maps of the sample surface, the local mechanical
properties of soft samples can be measured by application of very
small loading forces
\citep{Tao,Radmacher1995,Schaefer,Weisenhorn,Rotsch,Radmacher3}. By
use of the {\em Hertz model} for elastic indentations
\citep{Hertz,Sneddon}, Young's modulus $E$ can be determined.

Phospholipid membranes have extensively been studied by AFM
techniques. AFM was used to reveal information on the morphology and
topology of membranes, bilayers, domains
\citep{Perino,Giocondi,Dufrene,Kolb,Beckmann} and ripple phases,
\citep{Leidy} in real time by controlled environments
\citep{Kaasgaard,Tokumasu,Enders}. In addition, Fourier
Transformation was used to corroborate the periodicity of the ripple
structure \citep{Enders}. Single or double bilayers on solid
substrates as, e.g., silicon or glass, are usually used for these
studies with the drawback that the underlying substrate often
influences or evens governs the elastic response.

Multi lamellar stacks of these model membranes are often used
particularly in scattering experiments, to maximize the scattering
volume and improve signal to noise ratio considerably. Little is
known about the surface of the top layer in these stacks because the
membrane stacks are often not very stable but easily washed away
under excess water thereby making AFM investigations very difficult.
But this approach has the advantage that the interaction of the
probe and the underlying substrate can be neglected. Furthermore it
allows to quantify the forces to compress the stack and probe the
stability of the bilayers by determining the rupture limit. This can
be done at different temperatures, i.e., in the different phases of
the phospholipid membranes, the ripple phase (P$_{\beta'}$) and the
fluid L$_\alpha$ phase. This technique might therefore offer a new
and independent approach to the elasticity parameters, in particular
to the compressional modulus, $B$, of the stacked model membranes as
we will show below.

The elastic properties of lipid membranes, in particular in multi
lamellar stacks, have mainly been investigated by the analysis of
thermal diffuse scattering, i.e., x-ray lineshape analysis
\citep{Safinya:1986,Petrache:1998}. This work has led to a detailed
understanding of the static properties of thermal fluctuations in
lipid membranes and the elasticity properties governing these
fluctuations. A recent experiment \citep{RheinstaedterPRL:2006}
reported an inelastic neutron scattering study where the dispersion
relation of the mesoscopic fluctuation was determined by the neutron
spin echo technique. According to linear smectic elasticity theory
\citep{Caille:1972,Lei:1995} thermal fluctuations in the fluid phase
of the membranes are governed by the free energy functional
(Hamiltonian) \citep{Caille:1972,Lei:1995,LeiThesis:1993}
\begin{equation}
H = \int_A d^2r
\sum_{n=1}^{N-1}\left(\frac{1}{2}\frac{B}{d}(u_{n+1}-u_n)^2+\frac{1}{2}\kappa\left(\nabla^2_{||}u_n\right)^2\right)~,
\label{Hamiltonian}\end{equation} where $\kappa$ denotes the bilayer
bending rigidity, $A$ the area in the $xy$-plane, $N$ the number of
bilayers, and $u_n$ the deviation from the average position $(n\cdot
d)$ of the $n$-th bilayer, $d$ is the lamellar spacing. $B$ and $K =
\kappa/d$ are elastic coefficients, governing the compressional and
bending modes of the smectic phase, respectively. Note that the
surface tension $\sigma$ can usually be neglected because of the
boundary conditions, i.e., infinitely large membranes the rims of
which are not clamped and small amplitude of the fluctuations
\footnote{The deformation of the upper bilayer depends on the
intrinsic surface tension $\sigma$ and an elastic contribution from
the bulk smectic elasticity $(B\cdot K)^{1/2}$, in other words there
is an effective surface tension $\sigma _{eff} = \sigma + {\sqrt {K
B}}$ and for a more accurate estimate one has to compare $\sigma
_{eff} q_{||}^2$ and $\kappa q_{||}^4$ to neglect the surface
tension in the Hamiltonian.}.
$K$ and $B$ always appear as coupled parameters in scattering
experiments and only the use of aligned lipid bilayers allows a
separate determination of $K$ and $B$
\citep{Lyatskaya:2001,Salditt:2003}. In this work we used AFM to get
direct access to the compressional modulus $B$ of the membrane
stacks.


Our AFM experiments were motivated by the inelastic neutron
scattering study \citep{RheinstaedterPRL:2006}, which reported a
soft-mode in the dispersion relation of the long wavelength
undulation modes close to $T_m$. The critical $q$ value was found to
be $q_c\approx 0.015$\AA$^{-1}$ equivalent to a distance in real
space of $\lambda_c=2\pi/q_c\approx 420$~\AA. The relaxation rate of
the collective undulations dropped by more than two orders of
magnitude at $q_c$. A possible interpretation is that the well known
softening of phospholipid bilayers due to the decrease of the
bending modulus of the bilayers, $\kappa$ \cite{Chu:2005} occurs on
a particular length scale, only, similar to the freezing of a phonon
mode at elastic phase transitions in crystals. To clarify the origin
of this particular length scale we investigated the same samples
that have been used for the scattering experiments, namely thick
highly oriented lamellar membrane stacks on silicon wafers, in the
temperature resolved AFM experiments.

\section*{Materials and Methods\label{Materials and Methods}}

\subsection*{Atomic Force Microscopy}
A commercial AFM (Bioscope, Digital Instruments, Santa Barbara, CA)
combined with an Axiomat inverted optical microscope (Zeiss) was
used for this study. The optical microscope was used to position the
tip above the membranes. Soft silicon nitride cantilevers
(Microlever, Park Scientific, Sunnyvale, CA) with Gold reflex
coating on the cantilever back side and a spring constant of about
$6~mN/m$ were used. The spring constants were estimated from their
resonant frequency, which was $18~kHz$. The cantilevers had tip
semi-angles of $35~^{\circ}$. The wafer with the membrane probe was
magnetically attached to the microscope object stage. The microscope
was resting on a granite plate supported by soft rubber bands
attached to the ceiling for vibration isolation. Resistors glued to
the bottom of the object stage allowed to heat the object stage and
so the probe. A PT100 temperature sensor was mounted on the silicon
substrate and recorded the sample temperature during the
experiments. Because of the large thermal mass of the sample stage
the temperature stability was as good as $0.1~^{\circ}C$, the
absolute accuracy was better than $0.5~^{\circ}C$.

\subsection*{Samples}
Multilamellar samples composed of stacks of several thousands of
lipid bilayers separated by layers of water, resulting in a
structure of smectic A symmetry, have been prepared. DMPC
(1,2-dimyristoyl-sn-glycero-3-phoshatidylcholine) was obtained from
Avanti Polar Lipids. We then prepared highly oriented membrane
stacks by spreading a solution of typically $25~mg/ml$ lipid in
trifluoroethylene/chloroform (1:1) on 2'' silicon wafers
\citep{Muenster:1999}. The mosaicity of the samples, which is a
measure of the alignment with respect to the surface normal of the
stacked bilayers, had been determined by rocking scans to a few
hundreds of a degree. The samples have been carefully dried and
stored for several days in a refrigerator and after that in a
freezer. Note that in these conditions the membranes transform in a
densely packed sub-gel phase, which has been characterized only
recently \citep{Meyer:2000}. After warming to room temperature and
re-hydrating, the samples then abruptly irreversibly transform into
the ripple or fluid phase, respectively, and show the well known
reversible sequence of gel, ripple and fluid phase. The temperature
of the main phase transition for DMPC is at about $24~^{\circ}C$. In
samples, which had not been stored in freezers before, the membrane
film is easily washed away under excess water and no AFM
investigation was possible. We argue that the lipid film is
stabilized when the border areas, which are not in contact with
water or solution, are still in the very stable and dry sub-gel
phase. The effect of osmotic pressure has been studied by measuring
solutions of (Millipore) water with different concentrations of PEG
(polyethylene glycol with a molecular weight of 20000). About $2~ml$
of highly pure Millipore water or Millipore/PEG solution was dropped
onto the sample for the measurements, covering a sample area of
about ($\pi$ (0.5 cm)$^2$), and the AFM cantilever arm was immersed
into the water.

\subsection*{Data Acquisition and Analysis\label{Acquisition}}
Using an AFM, the surface is scanned by a nanometer sized tip, which
is mounted on an elastic cantilever arm. The size of the tip is in
the order of about $40~nm$. The deformation of the cantilever arm is
measured and the force constant of the cantilever arms then allows
to determine the force (attractive or repulsive) between sample and
AFM probe. By defining a threshold, the {\em contact} point between
sample and tip, and with it the height profile of the sample can be
determined. A second mode of operation is to continuously scan the
tip over the sample surface and continuously measure the deflection
of the cantilever arm. The result is usually displayed in two
dimensional map plots.

\begin{figure}[p] \centering
\includegraphics*[width=3.25in]{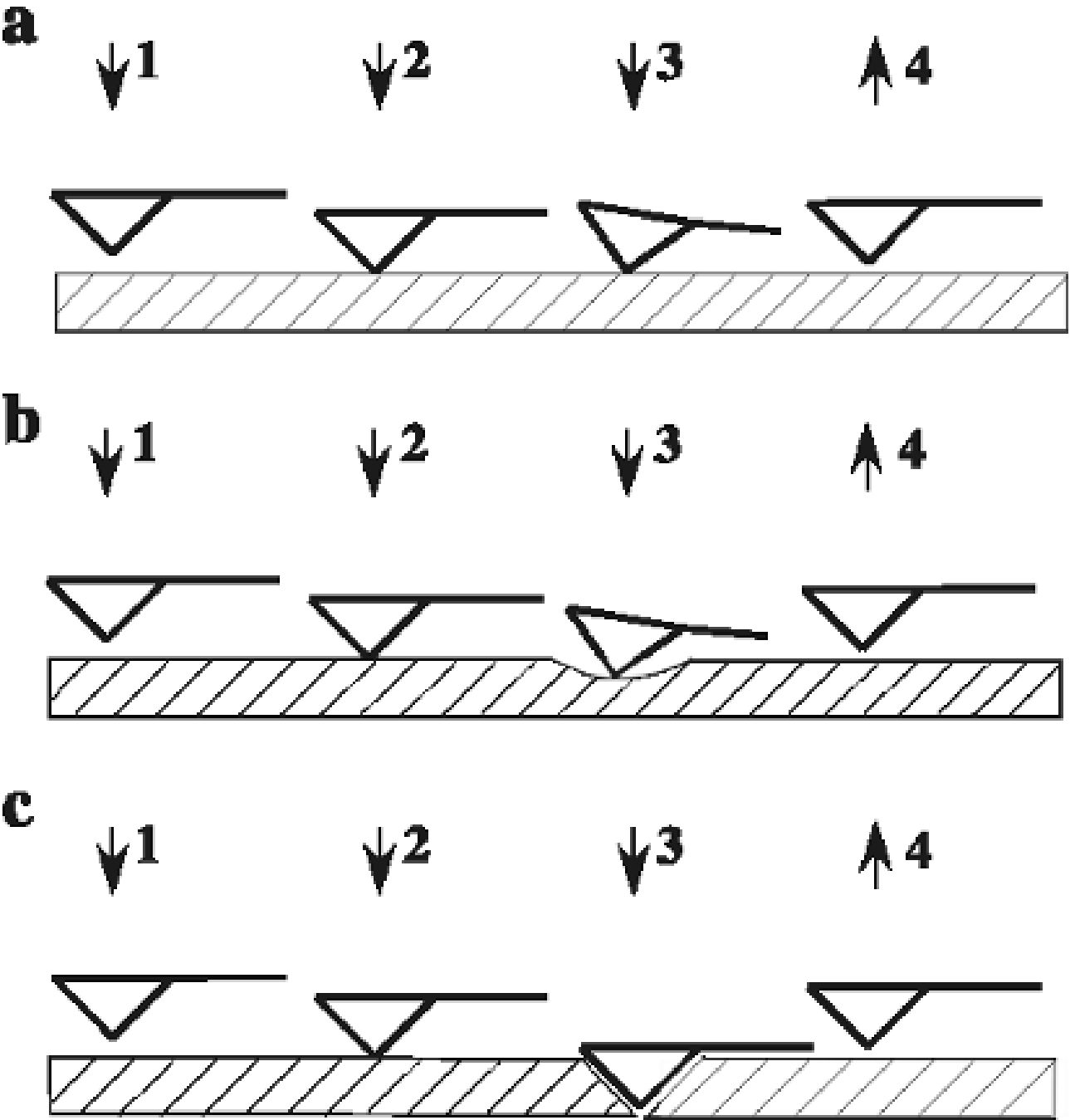}
\caption{
} \label{Skizze_2a}
\end{figure}

\begin{figure}[p] \centering
\includegraphics*[width=3.25in]{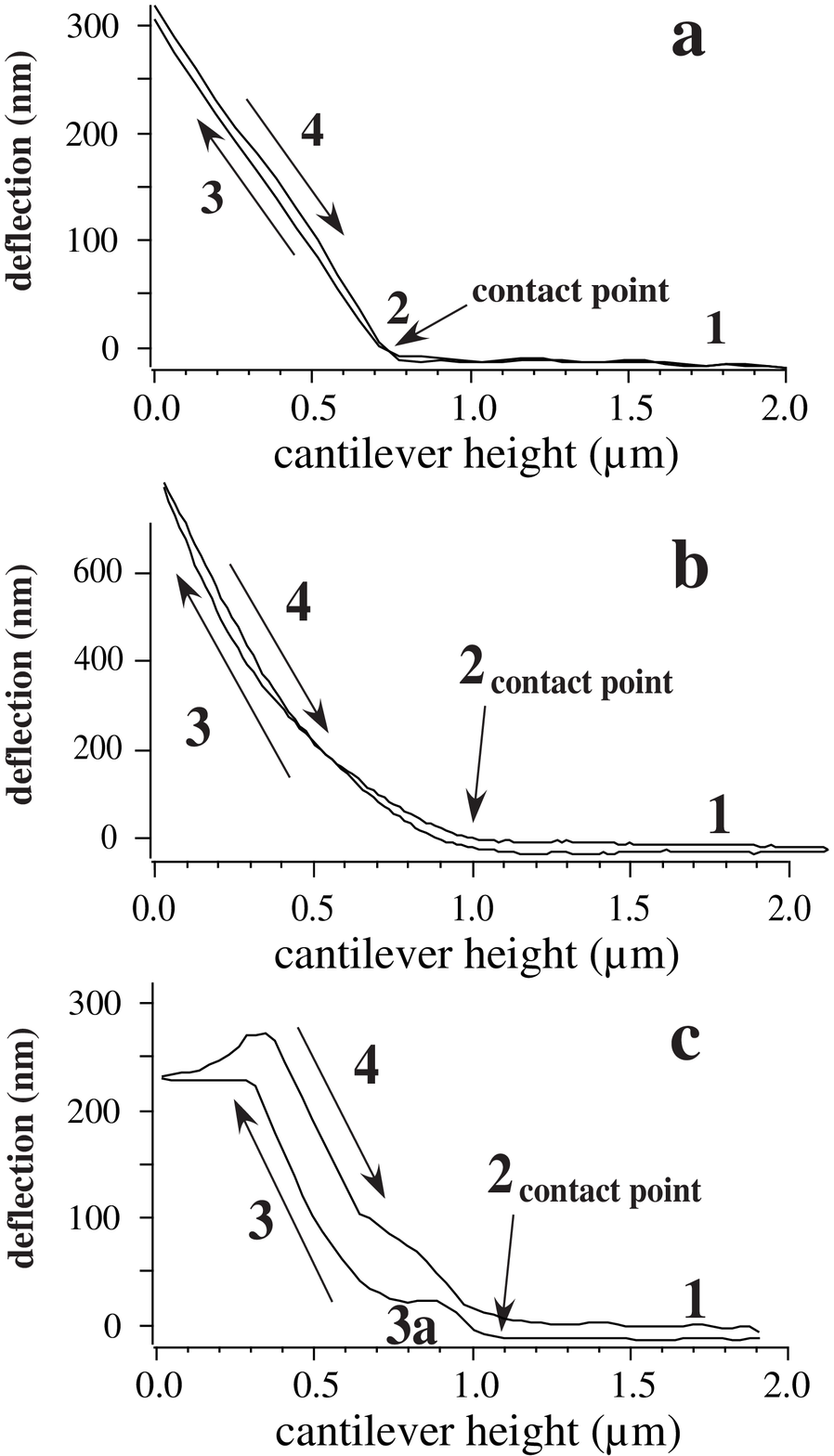}
\caption{
} \label{Force}
\end{figure}

Different scenarii for the interaction and the elastic response
between tip and sample can be discussed and are exemplary pictured
in Fig.~\ref{Skizze_2a}. When the tip approaches and comes in
contact with an infinite stiff surface in Fig.~\ref{Skizze_2a} (a)
(a dry membrane stack) the cantilever arm will start to bend after
being in contact. The elastic response in this case is perfectly
linear and determined by the spring constant $k_c$ of the cantilever
arm. If the surface is soft or not robust (in the case of a hydrated
membrane stack), the tip does indent or cut into the surface and
there is few or no bending of the arm when lowering the cantilever,
see Fig.~\ref{Skizze_2a} (c). In general, a surface will show a
mixed effect. The surface will be incised and if the force becomes
too high, the tip might cut into the surface, as depicted in
Fig.~\ref{Skizze_2a} (b). In the corresponding force curves, which
are shown in Fig.~\ref{Force} (a), (b) and (c) for the three cases
discussed above, the deflection of the AFM cantilever is monitored
as a function of its vertical position while approaching the sample.
While there is a linear slope of the curve when deflected from a
rigid surface in Fig.~\ref{Force} (a), the soft sample in
Fig.~\ref{Force} (b) is incised when lowering the AFM tip. In
Fig.~\ref{Force} (c), the tip is first deflected (points 1-3) but
then cuts through the surface as indicated by the horizontal
plateau. From the slope of these curves, normalized to an infinite
stiff surface, the forces to press and cut through the membranes can
be quantified \citep{Schaefer}.

\subsection*{Geometry\label{Geometry}}
The elastic parameters of the bilayers which are actually probed by
the AFM
depend on the particular experimental geometry. 
The elasticity of lipid membranes spanned over small $nm$ sized
pores, where the tip size is in the order of the size of the
membrane with correspondingly large indentations, has been probed by
AFM very recently \citep{Steltenkamp:2006}. The theoretical
framework that describes the indentation of the pore-spanning
bilayers has been developed and published in parallel
\citep{Norouzi:2006} and the indentation profile of a parabolic tip
and the bilayer has been carefully investigated. Note that this
setup leads to strong curvature and non-linear equations. For pore
spanning membranes on large holes (tip size and indentations small
as compared to the size of the membrane) the assumption of small
displacements is reasonable and the Young's modulus $E$ of the
membranes is probed. AFM investigations on single supported
membranes allow to study breakthrough forces, breakthrough
distances, adhesion, stiffness and topography
\citep{Kunnecke:2004,KunneckeCPC:2004} with a high spatial
resolution.

The present work deals with very thick multi-lamellar stacks of
solid supported membranes. Applying force onto membrane stacks is an
entirely different situation than the case addressed above on single
membranes. The pressure applied by the tip will lead to a local
compression of the membrane stack, leading to a decrease of the
inter-bilayer spacing $d$ (lamellar repeat distance) from the
equilibrium one. Note also, that in contrast to a uniform linear
force response, some force curves exhibit horizontal plateaus, i.e.,
the tip lowers without deflection, indicative for a cut through the
bilayers. In the linear regime, a compression of the stack will
result over some lateral interaction area $A$. Apart from
experimental parameters like tip radius $A$ will be determined by
the interplay of $B$ and $K$ as well as the swelling state $d$.
According to Eq.~(\ref{Hamiltonian}), a local force will lead to a
compression over some interaction area $A$, to minimize bending and
compressional energy. If for an idealized  patch of area $A$ we
treat the membrane compression, in an oversimplistic approach, as a
compression of flat and stiff planes, the compressional modulus B of
the membrane stack can be written as the derivative of the applied
force with respect to the cantilever height, normalized to the
lamellar $d$ spacing of the bilayers

\begin{equation} B = \frac{d}{A} ~\frac{\partial F}{\partial z}.
\label{BEqu}\end{equation}

The entire physics of smectic elasticity is now contained in the
effective parameter $A$. The higher the bending rigidity and the
lower the compressional modulus, the larger $A$.  The basic length
scale of this interplay is the smectic length scale $\lambda_c =
\sqrt{K/B}$. We thus postulate in $A\propto \lambda_c^2 = K/B$, with
$\lambda_c$ typically on the same order of magnitude, but somewhat
smaller than the interlamellar spacing $d$.

The approach is justified a posteriori by the values that are
obtained for $B$, if typical values for $\lambda_c$ are taken to
estimate $A$. The values for $B$ then agree well with values
reported in the literature by other techniques, see below. Note,
however, that it is the bending modulus $K=d \frac{\partial
F}{\partial z}$ which determines the slope of the force curve, since
$B$ cancels out in Eq.~(\ref{BEqu}).

\section*{Results and Discussion\label{Results}}
\subsection*{Force Curves\label{ForceCurves}}
Force curves have been measured at different temperatures. The
sample has been checked by deflection scans and an area clean from
obvious defects like steps, holes or kinks has been selected.
Typical deflection curves as a function of cantilever height are
shown in Fig.~\ref{Kraftkurven.EPSF} (a). The dry bilayer stack at
$19~^{\circ}C$ has been measured as a reference. It shows the
characteristics of a rigid sample with a well defined contact point
and well defined linear slope, as discussed in Figs.~\ref{Skizze_2a}
and \ref{Force}. After applying the excess water the response
changes and the surface becomes softer with a less well defined
contact, but still linear elastic response. The situation is then
completely different in the fluid phase at, e.g.,
$T=26.7~^{\circ}C$. 
When approaching, the tip is first deflected, but then cuts through
the first layers, as indicated by the horizontal plateau, before it
is deflected again. There is a critical deflection above which the
tip is just cutting through the bilayers. Although the positions of
the plateaus might slightly vary in different experiments, we found
this as typical response in the fluid phase.

Close to the phase transition, at $T=23.4~^{\circ}C$, in the range
of the critical swelling of the phospholipid bilayers, the membranes
are more stable, there is no rupture of the bilayers, but initially
there is only little deflection when lowering the tip indicating
that the stack is very soft and can easily be compressed. The slope
of the curve becomes steeper at higher indentations, only. From
these data, $B$ and the rupture force $F_r$ can be determined. 

Taking the dry sample as an ideal stiff reference, we can quantify
the corresponding interaction forces. 
The difference from using, e.g., the silicon substrate as stiff
reference is that the dry-membrane could be measured in-situ thereby
preserving the interaction between tip and sample and also the
geometrical set-up. Note that the normalization to the dried
membrane stack cancels out all elasticity already present without
excess water at ambient conditions (room temperature and humidity).
The force is then calculated to $F=k_c(z-z_{ref})$, $k_c$ is the
cantilever spring constant, $(z-z_{ref})$ the difference of
cantilever height with respect to the (hard) reference sample.
Figure~\ref{Kraftkurven.EPSF} (b) plots the resulting forces in $nN$
versus the cantilever height $z$ for the different temperatures.

$B$ has been determined from the slope of the force curves in
Fig.~\ref{Kraftkurven.EPSF} (b) following Eq.~(\ref{BEqu}), and is
plotted in Fig.~\ref{Bmodule} for $T=19.0~^{\circ}C$,
$T=23.4~^{\circ}C$ and $T=26.7~^{\circ}C$, i.e., in the gel (ripple)
phase of the bilayers, close to the phase transition and in the
fluid phase. The corresponding lamellar spacings $d$ can of course
not be determined by AFM but have additionally been measured by
x-ray diffraction on multi lamellar DMPC vesicles in excess water.
Vesicles have been prepared and used to ensure full hydration of the
bilayers because the hydration can be expected to be a crucial
parameter for the compressibility of the membrane stacks. While the
elastic response of the membrane stack is linear in the gel phase as
indicated by the constant value for $B$ in Fig.~\ref{Bmodule} at
$T=19.0~^{\circ}C$, elastic responses in the range of critical
swelling and in the fluid phase are highly nonlinear.
Tab.~\ref{Bmodulus} lists values for the rupture force $F_r$ and
$B$, as determined from Figs.~\ref{Kraftkurven.EPSF} and
\ref{Bmodule}. In the nonlinear regimes, the values for $B$ and
$F_r$ are the maxima and minima, respectively, of all the values
that we measured.

In the gel phase, we find $B$ values of 4.4$\cdot 10^6N/m^2$. Close
to $T_m$, the compressional modulus more than doubles to $B>
8.3\cdot 10^6N/m^2$. Although the corresponding curve in
Fig.~\ref{Kraftkurven.EPSF} (b) looks indeed much {\em softer} than
in the gel phase, the slope at higher indentations is actually much
steeper. The slope of the force-distance curve in the fluid phase is
again smaller than in the gel phase and $B$ is determined to
$B>3.3\cdot 10^6N/m^2$ (interrupted by the bilayer rupture). All
values for $B$ agree quite well with and lie quite in the middle of
the broad range of values reported in the literature
\citep{Liu:2004,Pabst:2003,Chu:2005,Petrache:1998}. While the energy
to compress the membrane stack is comparable in gel (ripple) and
fluid phase, the stack is distinctly harder to compress in the range
of {\em critical swelling}. We thus argue that AFM offers an
independent determination of the compressional modulus of membrane
stacks. While in scattering experiments it is usually very difficult
to determine the bending modulus $K$ and $B$ independently from each
other, the determination of force curves by AFM gives direct access
to $B$. Although that our approach might be oversimplified, it
nevertheless gives very reasonable values for the compressional
modulus of the membrane stacks. The dynamic experiment in
\citep{RheinstaedterPRL:2006} gave a larger $B$ of $B=1.08\cdot
10^7N/m^2$ in the fluid phase at $30~^{\circ}C$, which might be
related to only partial hydration of the membranes in the neutron
experiment. The bilayers were hydrated from the vapor phase and
swollen to $d=54$ \AA, only

\subsection*{Ripples on DMPC\label{Ripples}}
The preparation of thick membrane stacks also allowed to study the
structure of the top bilayer and to determine the ripple periodicity
$\Lambda_r$. Starting at a temperature of $18~^{\circ}C$, the sample
was slowly heated with a rate of $0.1~^{\circ}C$/min and deflection
was continuously scanned over an area of $1~\mu$m$\times$$1~\mu$m.
Static ripple patterns have been observed. To determine the
corresponding ripple period $\Lambda_r$, we have selected an area of
$300~nm\times 300~nm$ and Fourier transformed. From the sharp spots,
the corresponding $\vec{k}$-vector and $\Lambda_r=1/|\vec{k}|$ was
determined. The resulting plots are shown in
Fig.~\ref{03012006_Bilder.eps}.
 Figure~\ref{03012006.eps} gives the resulting period for all
measured temperatures. Note that there is a certain spread in the
determination of $\Lambda_r$. The Fourier transform integrates over
an area of $300~nm\times 300~nm$ and the occurrence of different
ripple domains, i.e., different directions of the normal vectors,
and static defects might lead to slightly different $|\vec{k}|$
values and periods.

From the temperature dependence of $\Lambda_r$ in
Fig.~\ref{03012006.eps}, the ripple period is almost constant with
values of about 100~\AA\ in the ripple phase at temperatures
$T<T_m$. $\Lambda_r$ increases when approaching the main transition
at about $24~^{\circ}C$. At temperatures in the range of T$_m$ we
find a coexistence of ripples with $\Lambda_r$ and $2\Lambda_r$
(about 420 \AA), visible by two diffraction spots in the Fourier
transformation in Fig.~\ref{03012006_Bilder.eps}.
Figure~\ref{03012006.eps} suggests a diverging ripple period when
approaching T$_m$. Weak ripples stay visible in the Fourier spectra
up to about $T=25~^{\circ}C$, 1 degree above $T_m$. At higher
temperatures in the fluid phase, no ripples can be detected.

Metastable $2\Lambda_r$ and even $4\Lambda_r$ ripples have been
reported earlier from AFM experiments on double bilayers
\citep{Kaasgaard,Fang:1996} and x-ray diffraction
\citep{Katsaras:2000}. The length scale of the metastable
$2\Lambda_r$ ripples agrees quite well with the length scale of the
soft-mode found in the dynamical neutron study with
$2\Lambda_r\approx 2\pi/q_c$. We speculate that the softening in the
range of the phase transition might be coupled to the occurrence of
these metastable $2\Lambda_r$ ripples. A possible explanation is
that bending of the bilayers might occur mainly in the interfaces
between two metastable ripples. The bending modulus might thus be
distinctly softer because structure and interactions are likely to
be much less well defined in the interface and not much energy
needed to slightly change the tilt angle between to ripple flanks.
From our experiments we can not determine the structure of the super
ripple structure, i.e., if possibly a pattern of coexisting gel and
fluid domains develops within the super ripples structure, as
suggested by Heimburg \citep{Heimburg:2000}. It is well known that
large fluid and gel domains with $\mu m$ sizes coexist in the range
of the main transition which can for instance be observed by
fluorescence microscopy techniques. The inelastic neutron scattering
study in \citep{RheinstaedterPRL:2006} and our AFM study now point
to coexisting {\em nanodomains} with sizes of less than $50~nm$,
which might play a crucial role for the understanding of
phospholipid bilayer dynamics and criticality in the range of $T_m$,
the range of {\em critical swelling}
\citep{Chen:1997,Nagle:1998,Mason:2001}. Interestingly, the
existence of coexisting gel and fluid nanodomains close to $T_m$ has
been argued in a preceding inelastic neutron scattering study
\citep{RheinstaedterPRL:2004} to investigate the collective short
wavelength dynamics in DMPC bilayers, from the coexistence of gel
and fluid like excitations in the range of the critical swelling.
The existence of coexisting small gel and fluid domains has also
been argued by preceding AFM investigations
\citep{Xie:2002,Tokumasu} to compensate the large stress which
occurs at $T_m$ due to the volume difference of the two phases. Note
that \citep{Fang:1996} also reported metastable $2\Lambda_r$ ripples
which were distinctly 'softer' than the
$\Lambda_r$ ripples.

\subsection*{Ripple period on DMPC \& PEG\label{RipplesPEG}}
PEG exerts an osmotic pressure to the membrane stack which results
in a slightly reduced lamellar spacing $d$, as it is observed when
lowering the hydration of the stacked membranes. We have prepared
solutions of 5.8~\%, 12.1~\% and 25~\% PEG in Millipore water. About
$2~ml$ of the solution was then dropped onto the sample and the AFM
head immersed. All scans were taken at $T=19~^{\circ}C$, in the
ripple phase of the DMPC bilayers, where ripples could easily be
observed. For each concentration, several scans have been taken. The
resulting data are shown in Fig.~\ref{PEG}. Error bars might be
defined by the spread in $\Lambda_r$ in the different scans for the
same concentration. Although the error bars are relatively large,
there is a clear tendency: the ripple period $\Lambda_r$ increases
from about 100~\AA\ for DMPC under pure water to about 140~\AA\ for
DMPC in 25~\% PEG. Table~\ref{Pegtable} lists values for PEG
concentration, corresponding osmotic pressure and $d$ spacings of
the stacked membranes that can not be accessed by AFM but were taken
from the supplementary x-ray diffraction experiments. Using values
from Chu {\em et al.} \citep{Chu:2005}, the effect of an increasing
PEG concentration can be compared to lowering the hydration. The
corresponding values for RH are also given in Tab.~\ref{Pegtable}.

The osmotic pressure exerted by the polymer leads to an increase of
the ripple period $\Lambda_r$ while the lamellar spacing $d$
decreases. A correlation between ripple period and thickness of the
water layer, i.e., hydration has been established almost twenty
years ago \citep{Wack:1988} by x-ray diffraction of vesicles.
Although the technique was basically bulk sensitive and the samples
vesicles instead of highly oriented solid supported bilayers, the
authors found an increasing $\Lambda_r$ with decreasing lamellar
spacing $d$. According to an AFM study \citep{Fang:1996}, the
occurrence of ripples is a consequence of interbilayer interaction
and no ripples are observed on single bilayers. 
The increasing ripple period most likely stems from an increasing
interaction between the bilayers when decreasing the bilayer
distance. The ripples might for instance 'round out', i.e., the
amplitude and the tilt angle decrease, leading to an increased
period. But a structural investigation using x-ray or neutron
diffraction is clearly needed to determine the temperature dependent
ripple structure to draw further conclusions.

\section*{Conclusion\label{Conclusion}}
In conclusion we present an AFM study of highly aligned, solid
supported thick lamellar stack of phospholipid bilayers under excess
water and water/PEG solutions. Our investigation proves the
feasibility of working with AFM and thick samples in excess water.
These samples are often used to study structural and dynamic
properties of model membrane systems in x-ray and neutron scattering
experiments. An advantage as compared to single or double bilayers
is that in thick samples the interaction between the AFM tip and the
substrate can be neglected. From force curves at different
temperatures, we determined values for the compressional modulus,
$B$, and the rupture force, $F_r$, of the membrane stacks in the
ripple and the fluid L$_\alpha$ phase and in the regime of critical
swelling. The values for $B$ are in excellent agreement to
literature values and this novel approach allows to determine $B$
independently from other elasticity parameters.

The uppermost layer shows pronounced ripples in the P$_{\beta'}$
phase. We find an increasing ripple periodicity $\Lambda_r$ when
approaching the temperature of the main transition at
$T_m=24~^{\circ}C$. Close to $T_m$ we observe coexisting metastable
ripples with $2\Lambda_r\approx 420$ \AA. The length scale of these
ripples agrees well with the length scale of a soft mode in the
dispersion relation of the long wavelength undulation modes reported
in a recent inelastic neutron scattering study and might be
responsible for the well known softening of phospholipid bilayers in
the range of the critical swelling. The ripple period $\Lambda_r$
increases also with increasing osmotic pressure, most likely due to
an increasing interaction between the bilayers in the stack.

{\bf Acknowledgements:} We acknowledge financial support from the
DFG through project SA 772/8-2. MCR would like to thank the Institut
f\"ur R\"ontgenphysik, G\"ottingen, Germany, and particularly T.
Salditt for kind hospitality during his very enjoyable sabbatical
leave. We would like to thank E.~Kats for critically proof reading
and commenting on the manuscript in its final version.
\bibliography{Membranes_04092007}

\clearpage

\begin{table}\centering
\begin{tabular}{c|c|c|c}
T ($^{\circ}C$ & $d$ (\AA) & F$_r$ (nN) & $\frac{1}{d}\frac{\partial
F}{\partial z}$ $(10^6~N/m^2)$\\\hline
19.0 & 65.5 & $>$20 & ~~4.4\\
23.4 & 66 & $>$13 & $>$8.3\\
26.7 & 63 & $<$8 &  $>$3.9\end{tabular} \caption[]{Values for the
rupture force $F_r$, and the compressional modulus $B\approx
\frac{1}{d}\frac{\partial F}{\partial z}$ of the membrane stacks, as
determined from Figs.~\ref{Kraftkurven.EPSF} and \ref{Bmodule}. The
lamellar spacing $d$ has independently been determined by
supplementary x-ray diffraction experiments on multi lamellar DMPC
vesicles.}\label{Bmodulus}
\end{table}

\begin{table}\centering
\begin{tabular}{c|c|c|c|c|c}
\% PEG & log(P) ($dyn/cm^2$) & P ($10^5~Pa$) & $d$ (\AA$^{-1}$) & RH
(\%) & $\Lambda_r$ (\AA)\\\hline
0 & 0 & 0 & 65.5 & 100 & 100\\
5.8 & 5.55 & 0.35 & 60 & 99.98 & 105\\
12.1 & 6.08 & 1.2 & 58 & 99.925 & 120\\
25 & 6.94 & 8.7& 54 & 99.525 & 140
\end{tabular}
\caption[]{Osmotic pressure $P$ for the different PEG concentrations
 and corresponding $d$-spacing
\citep{Mason:2001,Petrache:1998,Mennicke:2006}. Values for the
relative humidity (RH) can be taken from, e.g., Fig.~3 in
\citep{Chu:2005}. $\Lambda_r$ periods have been taken as the average
values determined from Fig.~\ref{PEG}.}\label{Pegtable}
\end{table}

\clearpage
\section*{Figure Legends}
\subsubsection*{Figure~\ref{Skizze_2a}.}
Schematic of the AFM force measurement by indentation, for the case
of (a) a stiff surface, (b) a soft surface and (c) the cantilever
cuts the surface. Labels (1) to (4) in this graph refer to the
corresponding sequence in the force curves in Fig.\ref{Force}: (1)
The piezo is approaching to the surface. (2) The AFM tip is in
contact with the surface (contact point). (3) The piezo is moving
down, leading to the elastic indentation in the case of a soft
surface and maybe cuts through the surface. (4) The piezo is moving
up.

\subsubsection*{Figure~\ref{Force}.}
(a): Typical force curve on a dry, hard membrane stack. The
cantilever height is defined as the $z$ direction. The arrows mark
the contact points where the cantilever first comes in contact with
the sample surface. The approach and the retraction curves are also
indicated by arrows. The elastic response is linear and determined
by the spring constant $k_c$ of the cantilever (the slope of the
curve). (b): Typical Force curve on a soft, hydrated membrane stack.
The slope of this curve is considerably smaller than that of the
stiff sample. The difference results from the larger indentation of
the cantilever in the softer sample. (c): In this case, after
indenting the surface, the cantilever cuts through several layer
when the force exceeds a characteristic threshold marked by the
plateau at small cantilever heights (compare Fig.\ref{Skizze_2a}
c)). The cantilever has no 'resistance' and so no deflection can be
measured.

\subsubsection*{Figure~\ref{Kraftkurven.EPSF}.}
(a) Force curves at different temperatures and for a dry sample. (b)
Taking the dry sample as a reference, the corresponding force can be
quantified.

\subsubsection*{Figure~\ref{Bmodule}.}
B Modulus determined as $B\approx \frac{1}{d}\frac{\partial
F}{\partial z}$ (Eq.~(\ref{BEqu})) for $T=19.0~^{\circ}C$ for
$T=23.4~^{\circ}C$ and $T=26.7~^{\circ}C$, i.e., in the gel (ripple)
phase of the bilayers, close to the phase transition and in the
fluid phase. While the elastic response of the membrane stack is
linear in the gel phase as indicated by the constant value for
$\frac{1}{d}\frac{\partial F}{\partial z}$ at $T=19.0~^{\circ}C$,
elastic responses in the range of critical swelling and in the fluid
phase are highly nonlinear. Tab.~\ref{Bmodulus} lists values for the
rupture force $F_r$ and $B$.

\subsubsection*{Figure~\ref{03012006_Bilder.eps}.}
AFM deflection scans over areas of $300~nm\times 300~nm$ and the
corresponding Fourier transform for selected temperatures: a)
$21.2~^{\circ}C$, b) $22.2~^{\circ}C$, c) $23.3~^{\circ}C$, d)
$23.9~^{\circ}C$, e) \& f) $24.2~^{\circ}C$. The arrows mark the
positions of diffraction spots in the Fourier spectra. The
corresponding period in real space is indicated by the lines in the
deflection plots. The ripple period $\Lambda_r$ increases when
approaching $T_m$ (k decreases) and at the same time the ripples can
be less well resolved, most likely due to a decreasing ripple
amplitude. e) \& f) Metastable $2\Lambda_r$ ripples (denoted ${\bf
k'}$) appear at the transition temperature which coexist with the
$\Lambda_r$ ripples (${\bf k}$). Ripple amplitude is very small at
$T_m$ so that the Fourier transformation is needed to unambiguously
determine $\Lambda_r$.

\subsubsection*{Figure~\ref{03012006.eps}.}
Temperature dependence of the ripple period $\Lambda_r$ as
determined from the spots (Bragg reflections) in the Fourier
transforms in Fig.~\ref{03012006_Bilder.eps}. $\Lambda_r$ increases
when approaching the main transition at about $24~^{\circ}C$. At
temperatures close to T$_m$ there is a coexistence of ripples with
$\Lambda_r$ and $2\Lambda_r$. Solid line is a guide to the eye.

\subsubsection*{Figure~\ref{PEG}.}
Ripple period $\Lambda_r$ for different concentrations of PEG, i.e.,
for different osmotic pressures. Solind line is a linear fit.
Several measurements have been made at each concentration. The
errors bars are reasonably defined by the scattering of the
$\Lambda_r$ values which arise from integrating over several ripple
domains in the AFM measurements in Fig.~\ref{03012006_Bilder.eps}.
The average $\Lambda_r$ values increase with increasing osmotic
pressure from about 100 \AA\ to 140 \AA.

\clearpage

\clearpage

\clearpage
\begin{figure}[p] \centering
\includegraphics*[width=3.25in]{Fig_3.eps}
\caption[]{
}\label{Kraftkurven.EPSF}
\end{figure}

\clearpage
\begin{figure}[p]
\centering
\includegraphics*[width=3.25in]{Fig_4.eps}
\caption[]{
}\label{Bmodule}
\end{figure}

\clearpage
\begin{figure}[p] \centering
\includegraphics*[width=5in]{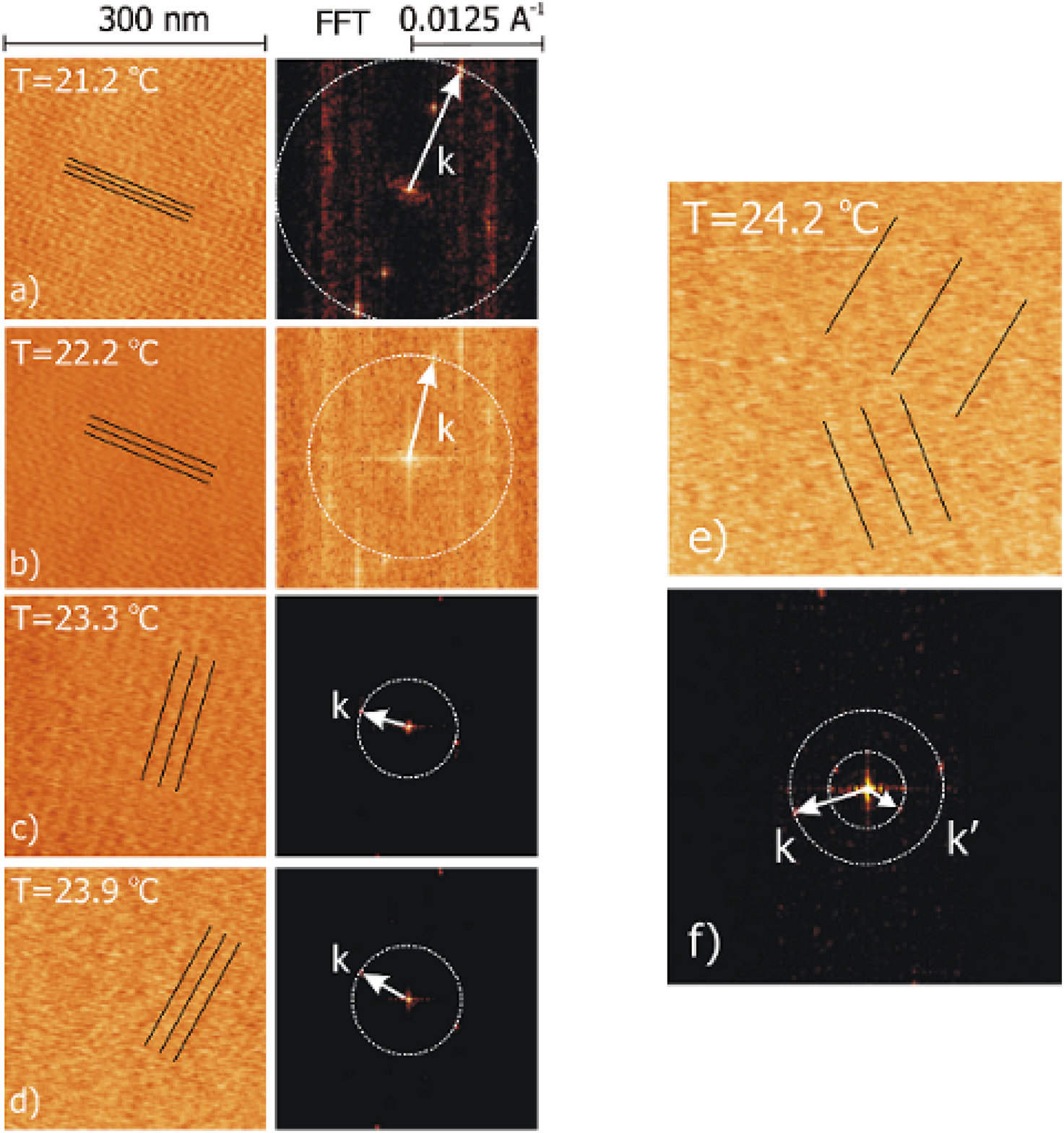}
\caption[]{
} \label{03012006_Bilder.eps}
\end{figure}

\clearpage
\begin{figure}[p] \centering
\includegraphics*[width=3.25in]{Fig_6.eps}
\caption[]{
}
\label{03012006.eps}
\end{figure}

\clearpage
\begin{figure}[p] \centering
\includegraphics*[width=3.25in]{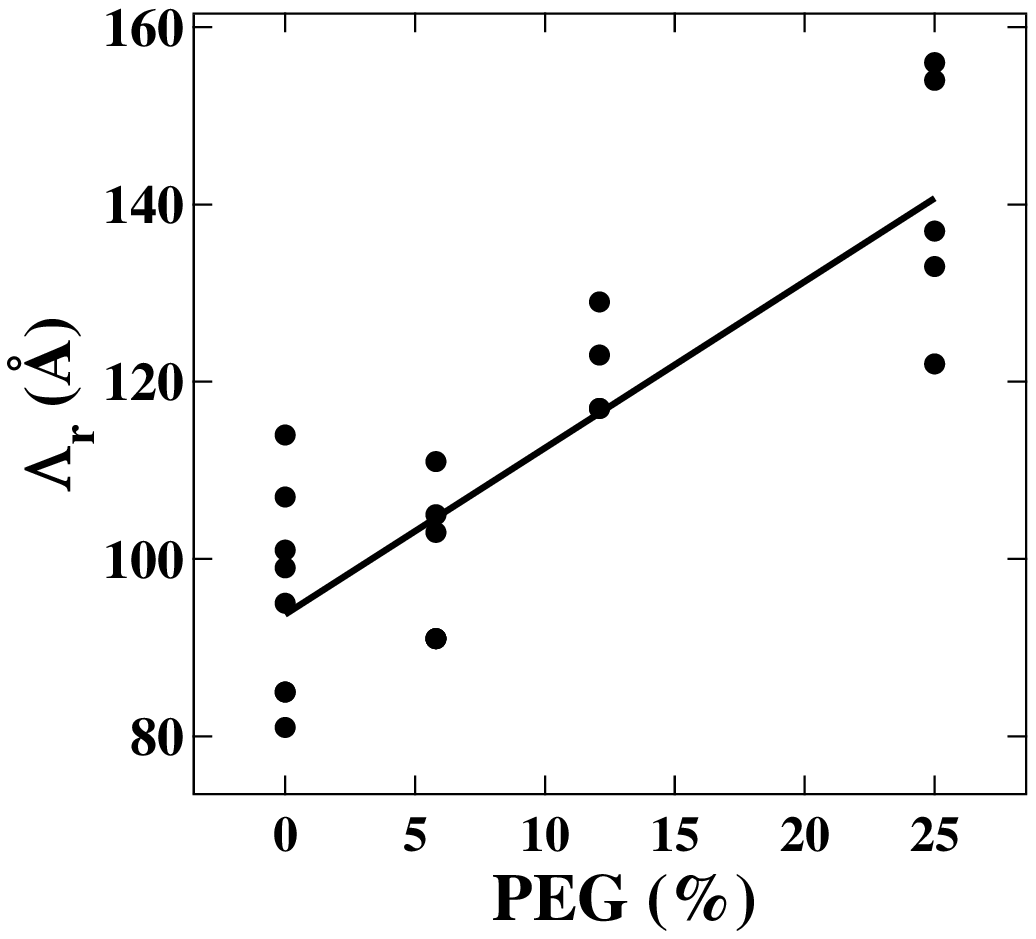}
\caption[]{
}\label{PEG}
\end{figure}


\end{document}